\documentclass[11pt,preprint,graphicx]{aastex}

\begin{document}

\title{Nonlinear Psychometric Thresholds for Physics and Mathematics}

\author{Stephen D.H. Hsu\altaffilmark{1} and 
James Schombert\altaffilmark{1}}

\altaffiltext{1}{Department of Physics, University of Oregon, Eugene, OR 97403}

\textbf{We analyze 5 years of student records at the University of Oregon to
estimate the probability of success (as defined by superior undergraduate
record; sufficient for admission to graduate school) in Physics and
Mathematics as a function of SAT-M score. We find evidence of a nonlinear
threshold: below SAT-M score of roughly 600, the probability of success is
very low. Interestingly, no similar threshold exists in other majors, such
as Sociology, History, English or Biology, whether on SAT combined, SAT-R
or SAT-M. Our findings have significant implications for the demographic
makeup of graduate populations in mathematically intensive subjects, given
the current distribution of SAT-M scores.}

\bigskip
\bigskip

A recent study ($1$) of 2000-2004 student records at the
University of Oregon (UO) showed correlations of 0.35 to 0.5 between SAT
scores and in-major, upper division GPA  (i.e., 300-400 level courses in 
the student's major). These correlations are higher than
those usually reported for SAT and freshman GPA ($2$,$3$)
most likely because
self-sorting of freshmen into more or less challenging courses weakens
correlations with ability measures. UO has a
number of favorable characteristics for this kind of study: it admits
students with a broad range of abilities (middle 50\% SAT-M + SAT-R: 990 -- 1220, so
little restriction of range), and the student body is relatively
homogeneous (thereby minimizing the impact of ethnicity).

Despite the somewhat higher correlations, we discovered impressive cases of
high achievement by students with relatively low SAT scores: in almost all
majors (e.g., English, History, Sociology, Biology, etc.) students with
combined (math + reading) scores well below 1000 (i.e., below the average
among all SAT-takers) achieved in-major, upper division GPAs (henceforth,
upper GPAs) in excess of 3.5 and even 4.0 (see figure 1 for a scatter plot
of upper division GPA in History and Sociology versus SAT-Reading). These
examples of the noisiness inherent in the SAT as a predictor of college
performance could be interpreted as cases where other factors, such as
extremely high conscientiousness or hard work, compensated for cognitive
limitations or limitations in college preparedness.
Overachievement of this type was found when comparing upper GPA to SAT
combined, or SAT-M or SAT-R individually. Indeed, the simplest mathematical
model one might formulate for GPA would include an ability input
(presumably normally distributed, and correlated with SAT) and another
input reflecting the effort exerted on the part of the student (this is
plausibly normally distributed; in our university-wide data the
fluctuations that remain after SAT is controlled for are roughly Gaussian
with fairly constant standard deviation).

However, two majors stood out as qualitatively different from the others.
In the cases of Physics and of pure Mathematics (defined by a set of
rigorous courses taken by graduate school bound majors), the pattern of
upper GPA versus SAT-M showed a sharp threshold: no student below
approximately SAT-M = 600 was able to attain the 3.5 upper GPA (i.e.,
roughly equal numbers of A and B grades) typically required for admission
to a Ph.D. program. (See figure 1, which displays upper GPAs vs SAT-M for
Physics and Mathematics graduates in our data set.) A small fraction of
students with SAT-M score at or above 600 attained upper GPA $> 3.5$; it
seems plausible that these were the most conscientious and determined of
students at this ability level. The fraction of students with high upper
GPA increases monotonically with SAT-M. Note that upper GPA $> 3.5$ is not
a high bar -- the average of all upper division grades in these departments
is about $3.2 \pm .6$, so $3.5$ is only half a standard deviation
above the average. The selection of GPA of 3.5 as a minimum value for
admission to graduate school is, of course, somewhat arbitrary, but
reflects our experience both with admissions to our graduate program in
Physics and with the subsequent career paths of undergraduate majors at
Oregon.

\begin{figure}
\centering
\includegraphics[scale=0.8]{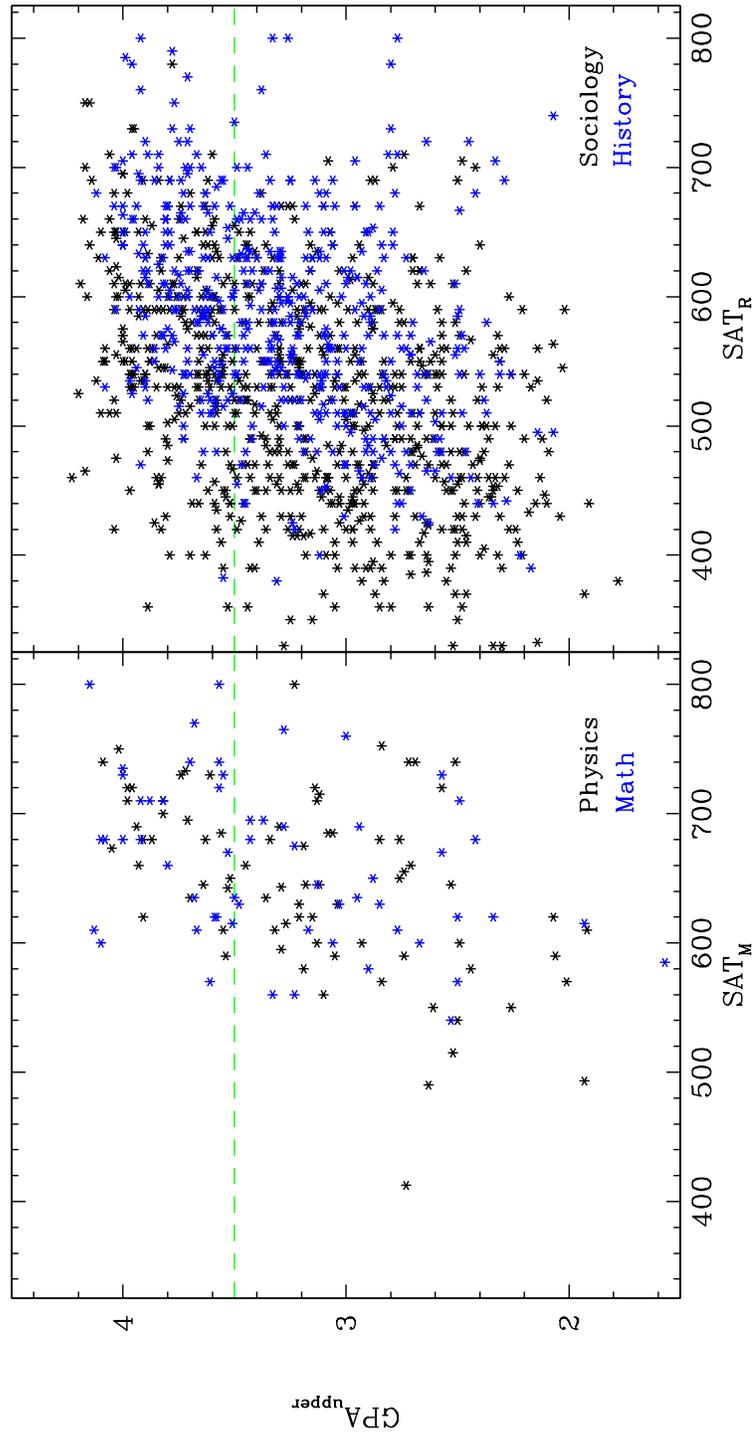}
\caption{Modest evidence for a cognitive threshold is found in the SAT Math
scores of Physics and Mathematics majors (left panel), but is not seen in
SAT (Reading or otherwise) scores for majors such as Sociology and History
(right panel). The green line indicates GPA of 3.5.}
\end{figure}

The existence of a minimum threshold, measurable by standardized tests,
required for success in Mathematics and Physics has numerous important
consequences. We expect that similar results also apply to highly
mathematical fields of study such as some areas engineering or informatics.

1. If one regards SAT-M as simply a diagnostic, the score of 600 becomes a
useful measure of {\it readiness} for college work in these fields. Indeed,
our data suggests that the probability of success below this level is very
low (see analysis to follow). Students who test below this level and who
intend to major in a mathematically intensive subject should be offered
enrichment in the basics of high school level mathematics before continuing
on to college level work in these fields. Programs meant to increase the
presence of underrepresented groups in STEM fields should focus on basic
skills -- perhaps a gap year after high school which focuses on raising
SAT-M related capabilities would prove successful. Many universities offer
intensive summer preparation programs in basic college skills for incoming students
from underrepresented groups. It is possible that prospective STEM majors with
low SAT-M scores would benefit from similar programs specifically focused on 
math skills.

2. If one regards the SAT as a measure of general cognitive ability ($4$,$5$),
then the 600 minimum gives an indication of the
intrinsic difficulty level of the university curriculum in Physics and
Mathematics. As such it has important psychometric implications. SAT-M 600
is roughly 75th percentile among test takers, and, depending on assumptions
concerning the SAT-taking population relative to the whole, roughly 85th
percentile in the general population. Thus, a strong interpretation of our
result would be that even the most determined student is unlikely to master
undergraduate Physics or Mathematics if their quantitative ability is below
85th percentile in the overall population. To have a 50 percent or greater
chance of success (i.e., for a person of average conscientiousness or work
ethic), one needs SAT-M well above 700, or in the top few percent of the
overall population.

Under {\it either} interpretation 1. or 2., a minimum threshold of 600 has
important implications. Because 600 is the {\it lowest} score at which even
the most determined students can succeed, the bulk of students in Ph.D.
programs in these fields will tend to have SAT-M scores of 650 or even
$700+$. This estimate is consistent with average GRE-Q scores well above
700 for applicants to programs in physics, math, engineering and computer
science (see supplementary materials of Ceci, Williams \& Barnett 2009
($6$),
or http://www.ets.org/Media/Tests/GRE/pdf/gre\_0910\_guide.pdf).  In our
UO data, same student SAT-M and GRE-Q scores exhibit a correlation of 0.75
and best-fit linear slope of close to unity. Therefore, the average
graduate school applicant in these fields likely scored well above 700 on
SAT-M.  At SAT-M score of 650, the ratio of white males to females is
almost 2:1, the percentage ratio of white males to African-Americans is
roughly 10:1, and the percentage ratio of Asian-Americans to white males is
almost 2:1. These ratios become even more extreme at higher scores -- see
College Board data on percentile equivalents by group at {\small
http://professionals.collegeboard.com/profdownload/SAT-Percentile-Ranks-by-Gender-Ethnicity-2009.pdf}.

Below we present evidence supporting the existence of a minimum threshold
at SAT-M roughly 600. Note, our analysis is agnostic as to possibilities 1.
and 2. above. The interpretation of the data in figure 1 suffers from the
relatively small numbers of students with low SAT-M. It is not hard to
understand this paucity: students with low SAT-M typically perform poorly
in Physics and Mathematics, and are unlikely to persist in the major
through graduation. Our records show that many students begin the physics
sequence, but later switch majors. In the case of Mathematics, there are
several tracks through the major and only a small fraction of students
attempt the rigorous courses (necessary for graduate study) we used to
calculate upper GPA. The freedom of students to choose their major and to
choose their curriculum makes it difficult to study the implications of low
SAT-M for success in these fields.

To remedy this problem we consider student records during the same
2000-2004 period for the year long calculus-based physics course
(251-252-253, henceforth 25X; UO is on the quarter system) taken by all
Physics majors and many Mathematics and Chemistry majors. 25X is not an
upper division course, but it has a standard and relatively rigorous
syllabus. We have verified that the ``style'' of the course during this
period was relatively consistent. In total, we have 826 terms of grade
records for this course, and the associated SAT-M scores of the students.
The distribution of average grades in 25X is shown in figure 2. Note an
almost identical threshold, again at SAT-M just below 600.

\begin{figure}
\centering
\includegraphics[scale=0.8]{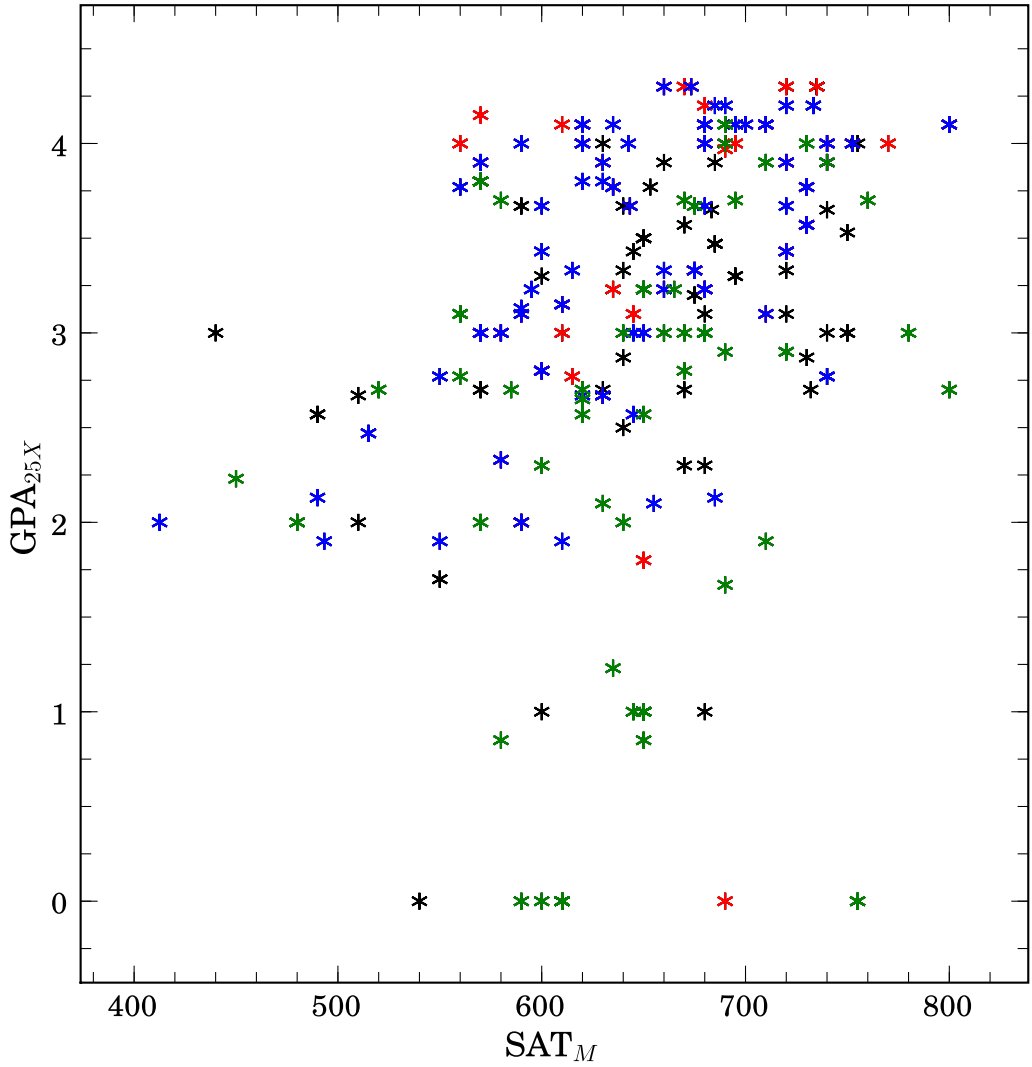}
\caption{SAT Math scores versus average GPA in PHYS 25X series of courses
(physics with calculus).  Blue symbols are Physics majors,
red are Math majors, black are other majors, green are students who did
not graduate from the University.}
\end{figure}

We further analyze the 25X data set as follows. Divide the students into
bins by SAT-M score, and estimate for each bin the probability $p$ that an
individual student earns an A-type (A+, A or A-) grade during particular
term.  (Bin sizes were varied in an effort to keep similar total numbers in
each bin, although this was not possible for the lowest and highest scoring bins.)
The simplest estimate of $p$ is obtained by computing the actual fraction of A
grades for students in the bin. However, we can do better than this using
the cumulative binomial distribution: we determine the largest $p$ which
is, at 95 percent confidence level, consistent with the number of A grades
actually earned. These $p$ values are displayed in figure 4 (red curve).
Note the increase in $p$ with SAT-M score.

\begin{figure}
\centering
\includegraphics[scale=0.8]{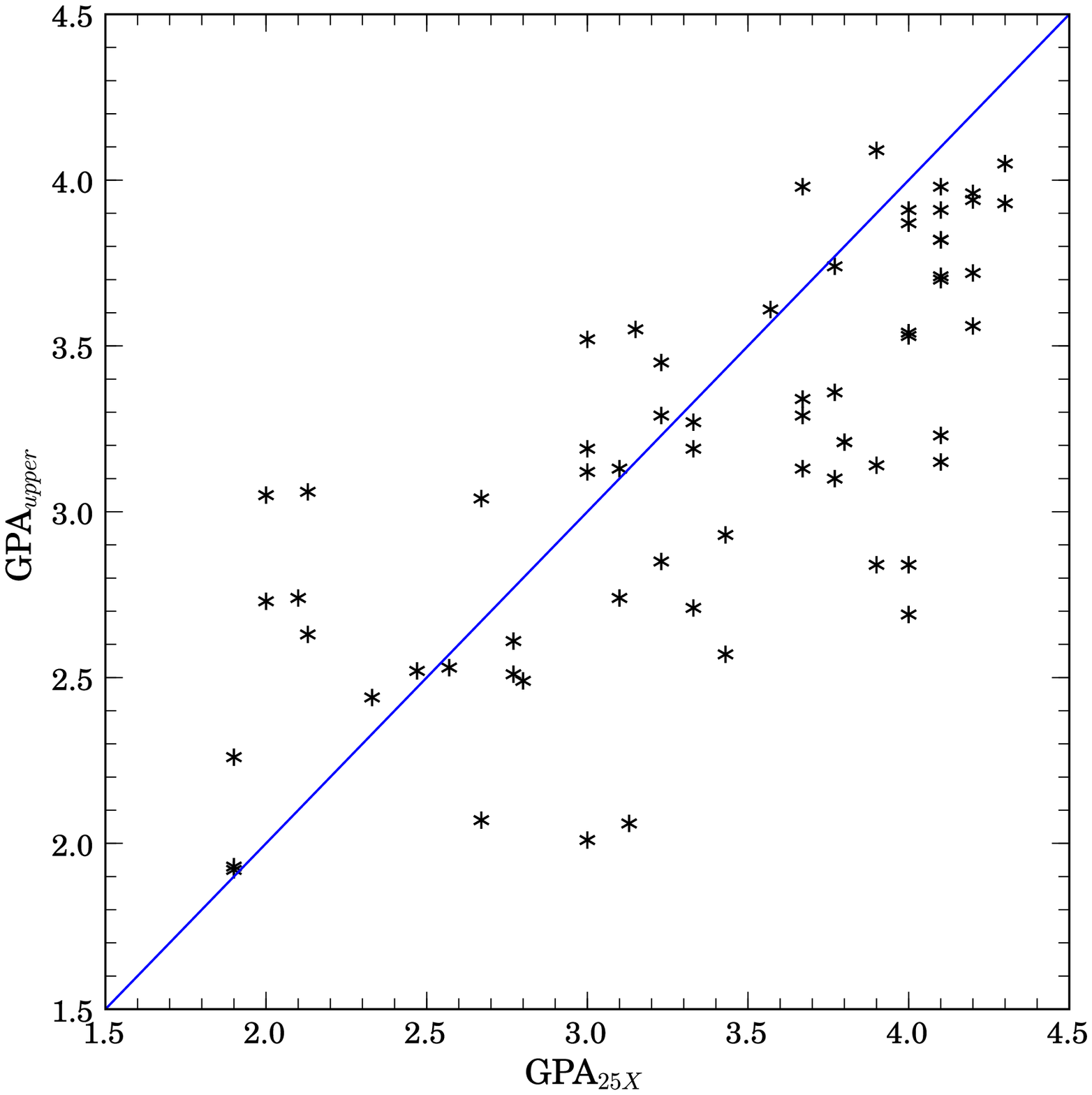}
\caption{The correlation between grades in PHYS 25X and upper division
Physics courses (300 or greater).  The blue line is equality.  The
correlation coefficient is 0.75.}
\end{figure}

We observe that upper division (300 and 400 level courses) are typically
much more challenging than 25X. Indeed, student grades in 25X correlate
0.75 with upper division GPA, with slightly higher grades earned, on
average, in 25X (see figure 3). Therefore, the probability $p_u$ of an
A-type grade in upper division courses is similar to but likely smaller
than the $p$ obtained above for 25X. We now derive an upper bound on the
probability $P( {\rm GPA} > 3.5)$ that a student in a particular SAT-M bin
can earn a cumulative upper GPA greater than 3.5. The typical graduate
school bound Physics major takes at least 16 upper division terms of
advanced Physics and Mathematics courses (again, we assume the quarter
system; another benefit of the Oregon data is the larger number of grade
records compared to another school on the semester system). To exceed upper
GPA  of 3.5 would typically require at least 8 A-type grades amongst these
16 terms (note A+ grades are relatively rare compared to A or A- grades).
Assuming $p > p_u$, we can deduce an upper bound on $P( {\rm GPA} > 3.5)$,
assuming that at least 8 A grades of some sort are required out of 16
courses in order to obtain an average upper GPA of 3.5: $P( {\rm GPA} >
3.5) \approx P( \geq 8 ~{\rm A's}~ | 16)$. The latter probability can be
calculated using $p_u$ and the binomial distribution. Results are shown in
the final column in figure 4: blue points are upper bounds on $P( {\rm GPA}
> 3.5)$ and the red curve is $p$ at 95 percent confidence level. We obtain
a threshold at around SAT-M 600: the probability of success (sufficient for
graduate school admission) is negligible at 95 percent confidence level for
SAT-M less than 600. For scores above 600 the success probabilities in the
table are increasingly large, but recall these are only {\it upper bounds},
for several reasons: we are using the largest single term success
probability $p$ that is compatible with data at 95 percent confidence; the
calculation essentially assumes that all non-A grades are some form of B,
which is optimistic, etc.

The error bars (figure 4) on the upper bound on $P( {\rm GPA} > 3.5)$ are
obtained using a Monte Carlo technique as follows.  In each SAT-M score bin,
we randomly remove half the data points and recompute $p$ and $P( {\rm GPA}
> 3.5)$. The errors shown indicate the variation in mean SAT-M score in
each bin (horizontal error bar) and in $P( {\rm GPA} > 3.5)$ (vertical
error bar), averaging over 100 such randomizations. We have checked that
these average variations are stable after 100 randomizations.

\begin{figure}
\centering
\includegraphics[scale=0.8]{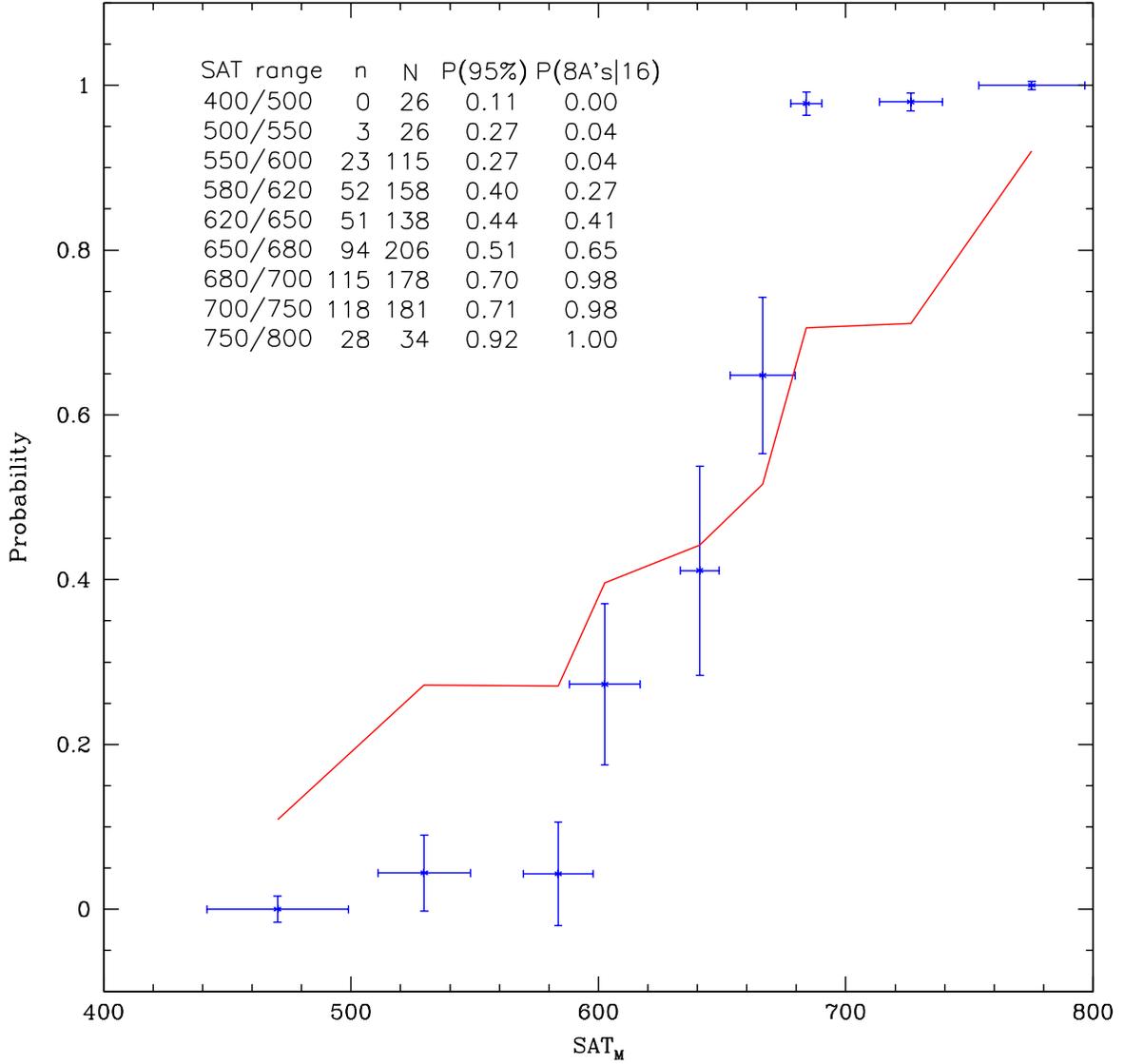}
\caption{Probability charts for SAT Math versus grades in Physics
classes.  For each SAT bin, the number of A-type grades (n) is listed
along with the total number of grades per bin (N).  The red line displays the 95\%
probability P(95\%) and the blue symbols display an upper bound on the
probability that a student will achieve 8 or more A's out of 16 courses
(typically necessary for an upper division GPA of 3.5.)}
\end{figure}

The upper bounds obtained in figure 4 are fully compatible with the actual
upper GPA data presented in figure 1, although note that even at very high
SAT-M score (e.g., well above 700) the actual observed fraction of students
who earn GPA $> 3.5$ never approaches unity. To have a fifty percent chance
of GPA $> 3.5$ probably requires SAT-M score well above 700 (i.e., in the
top few percent of the college bound population). It is plausible that for
a person of {\it average} conscientiousness or work effort to have a high
probability of performing well enough to be admitted to graduate school in
Physics or Mathematics requires math ability or readiness near the top
percentile.

We can give a simple, intuitive, explanation for the existence of the
observed threshold. While the probability of success (i.e., earning an A
grade) in any {\it single term} of Physics or Mathematics is smoothly varying and
nearly linear with SAT-M (see red curve in figure 4), the probability $P( \geq
8 ~{\rm A's}~ | 16)$ of earning 8 or more such A grades out of 16 courses
is highly non-linear in SAT-M. Once the single term success probability $p$
falls below a threshold, the cumulative GPA probability $P( \geq 8 ~{\rm
A's}~ | 16)$ drops to zero very fast. The non-linearity is generated by the
cumulative nature of the undergraduate grade record.

Throughout this discussion we have been agnostic to the underlying cause of
the SAT-M score-GPA correlation.  Under assumption 2. the SAT-M cut-off of
600 reflects general cognitive ability or $g$, which would require very
intensive intervention to elevate, in contrast to 1. which takes no
position on the underlying ability issue. Ultimately, this is an empirical
issue in need of actual data on the effect of various interventions to
elevate low-scoring students' performance.

\bigskip  

\noindent \textbf{References}

\noindent 1. Hsu, Stephen D. H. and Schombert,
James (2010). Data Mining the University: College GPA Predictions from Sat
Scores (April, 14 2010). Available at SSRN:
http://ssrn.com/abstract=1589792

\noindent 2. Sackett, Paul R., Borneman, Matthew 
J. \& Connelly, Brian S. American Psychologist. High Stakes Testing in 
Higher Education \& Employment, Vol 63(4), May-Jun 2008, 215-227. 

\noindent 3. Berry, C. M. \& Sackett, P. R. 
(2009). Individual differences in course choice result in underestimation 
of college admissions system validity. Psychological Science, 20, 822-830. 

\noindent 4. Park, G., Lubinski, D., and Benbow, C. P. 
(2008). Ability differences among people who have commensurate degrees 
matter for scientific creativity.  Psychological Science, 19, 957-961.

\noindent 5. Lubinski, D., and Benbow, C. P.
(2006). Study of Mathematically Precocious Youth after 35 years: Uncovering
antecedents for the development of math-science expertise. Perspectives on
Psychological Science, 1, 316-345.

\noindent 6. Ceci, S. J., Williams,
W.M., \& Barnett, S.M. (2009). Women's underrepresentation in science:
Sociocultural and biological considerations. Psychological Bulletin, 135,
218-261.

\end{document}